\documentclass[12pt]{iopart}

\usepackage{graphicx}
\usepackage{epstopdf}
\usepackage{subfigure}
\usepackage{verbatim}
\usepackage{gensymb}
\usepackage{url}
\usepackage{color}

\begin{document}

\title[]{Minimum energy path for the nucleation of misfit dislocations in Ge/Si(001) heteroepitaxy}

\author{O. Trushin$^1$, E. Maras$^{2,3}$, A. Stukowski$^4$, E. Granato$^{5,6}$, S.C. Ying$^6$, H. J\'onsson$^{2,7}$ and T. Ala-Nissila$^{2,3,6}$}

\address{$^1$ Institute of Physics and Technology, Yaroslavl Branch, Academy of Sciences of Russia, Yaroslavl 150007, Russia}
\address{$^2$ Department of Applied Physics, Aalto University School of Science, FIN-00076 Aalto, Espoo, Finland}
\address{$^3$ COMP Center of Excellence, Aalto University School of Science, FIN-00076 Aalto, Espoo, Finland}
\address{$^4$ Institut f\"ur Materialwissenschaft, Technische Universit\"at Darmstadt, D-64287 Darmstadt, Germany}
\address{$^5$ Laborat\'orio Associado de Sensores e Materiais, Instituto National de Pesquisas Espaciais, 12227-010 S\~ao Jos\'e dos Campos, SP Brasil}
\address{$^6$ Department of Physics, Box 1843, Brown University,
Providence, RI 02912-1843, U.S.A.}
\address{$^7$ Faculty of Physical Sciences, University of Iceland, 107 Reykjav\'{\i}k, Iceland}

\ead{otrushin@gmail.com}
\vspace{10pt}

\begin{abstract}
A 
possible
mechanism for the formation of a 90$\degree$  misfit dislocation at the Ge/Si(001) interface 
through homogeneous nucleation 
is identified
from atomic scale calculations 
where a
minimum energy path connecting the coherent epitaxial state 
and
a final state with a 90$\degree$ misfit dislocation is found using the nudged elastic band method.
The initial path is generated   
using a repulsive bias activation procedure 
in a model system including 75000 atoms.
The energy along the path exhibits 
two maxima in the energy.
The first maximum occurs as a 60$\degree$ dislocation nucleates. 
The intermediate minimum 
corresponds to an extended 60$\degree$ dislocation. 
The subsequent
energy maximum 
occurs as a second 60$\degree$ dislocation nucleates in a complementary, mirror glide plane, 
simultaneously starting from the surface and from the first 60$\degree$ dislocation. 
The activation energy of the nucleation of the second dislocation is 30\% lower than that of the first one showing that the formation of the second 
60$\degree$ dislocation is aided by the presence of the first one. 
The simulations represent
a step towards unraveling the formation mechanism of 90$\degree$ dislocations, an important issue in the design of growth procedures for strain released Ge overlayers on Si(100) surfaces,
and more generally illustrate an approach that can be used to 
gain insight into the mechanism of complex nucleation paths of extended defects in solids.
\end{abstract}
\maketitle
%
%
%
%
%

\section{Introduction}

Heteroepitaxial systems play 
an important role in modern microelectronics technology \cite{Campbell2001}.
Due to lattice mismatch between the
film and the substrate, considerable elastic strain energy is
accumulated during epitaxial growth. For sufficiently thick films, the strain energy
is released through the formation of defects, leading to the loss of
coherent epitaxy \cite{Luth1998,Bolkhovityanov2001}. Controlling the film-substrate interface quality
and preventing defect formation within
the film is an important 
goal
in modern technology. This requires detailed information 
about the complex atomic rearrangements during strain relaxation. 
Current experimental methods do not,
however, 
allow the measument of detailed atomistic
evolution in such processes.
Therefore, theoretical modeling 
can play an important role in studies of defect 
formation mechanisms and
atomic structure.

Misfit dislocations (MDs) are the most important type of defects in relaxed
heteroepitaxial systems \cite{Bean1985}. 
Early theoretical studies of MDs were based on the comparison of the energy of configurations with and without MDs \cite{Matthews1974,Ball1983}. Later on, it has
been recognized that the formation of defects leading to the loss of
coherent epitaxy is a thermally activated process, 
as illustrated by
the experimentally observed temperature dependence of
the critical thickness of the film \cite{Luth1998}. 
The state of the epitaxial system 
formed in experiments
is thus not only
determined by the nature and energy of the final state, but
also by the kinetic factors which depend on the 
transition path, and in particular on the energy barriers 
that need to be overcome to achieve
strain relaxation.

The Ge/Si heterostructure is a particularly important heteroepitaxial system
in microelectronics applications.
It is used in optoelectronics and for creating high speed electronic components \cite{Chaisakul2014,Liu2015,Bolkhovityanov2007}.
It is also an important substrate for GaAs growth on silicon \cite{Kim1997,Carlin2000}. 
The Ge lattice constant is around 4\% larger than that of Si, 
so a perfect, coherent film can only grow up to a
few monolayers. 
There are two competing mechanisms for relaxation of the film strain in Ge/Si(001), namely
through MD nucleation, and by 3D island formation in the Stranski-Krastanow growth mode.
Experiments show that the Stranski-Krastanow growth mode can be suppressed by proper choice of 
growth conditions (for example by using low temperature \cite{Eaglesham1991} and 
by using surfactants \cite{Bolkhovityanov2007}).
There are two main types of MDs in the diamond structure, namely  60$\degree$ and 90$\degree$ MDs. The terminology reflects the angle between the direction of the Burgers vector and the dislocation line. 
The nucleation of 60$\degree$ MDs in pure Si near a 
surface step under external stress has been extensively studied
 \cite{Shima2010,Li2010,Li2012,Izumi2008,Godet2004,Godet2006,Godet2009}. 
It has been
shown that since the 60$\degree$ MDs can glide on the
dense (111)
planes, they can form through half-loop nucleation on these planes.

The Burgers vector of a 90$\degree$ MD lies in the (001) plane while the Burgers vector of a 60$\degree$ MD does not. 
A 90$\degree$ MD
thus releases larger misfit strain than a 60$\degree$ MD. 
The 90$\degree$ MD, however, is sessile,
i.e. it is relatively immobile.
Experimentally,
a
regular array of edge MDs lying at the interface and 
almost entirely releasing
the strain from the film 
has been
obtained 
\cite{Liu2012,Bolkhovityanov2011}. Since 90$\degree$ MDs are sessile, 
the 90$\degree$ MDs must form through the reaction of non-sessile dislocations.
Bolkhovityanov \textit{et al.} have presented a review of 
various
possible
mechanisms for the formation of 90$\degree$ MDs \cite{Bolkhovityanov2011}. All the mechanisms described involve the reaction of two complementary 60$\degree$ MDs. Such a reaction could for example be of the form:
\begin{equation}\label{eqReac}
a/2[011](11\overline{1}) +a/2[10\overline{1}](111) \rightarrow a/2[110](001),
\end{equation}
where $a$ is the Si lattice constant, $a/2[011](11\overline{1})$ indicates that the MD has a Burgers vector of $a/2[011]$ and that it glides on the $(11\overline{1})$ plane.
In the induced nucleation mechanism, the presence of a 60$\degree$  MD favors the formation of a complementary 60$\degree$  MD on the mirror plane and 
the two
MDs react.
Two ways for the complementary MD to nucleate have been considered. The complementary MD can either form by a half-loop nucleation from the surface \cite{Matthews1975,Fitzgerald1991,Marzegalli2013} or it can form from the existing MD \cite{Gosling1993} following the reaction:
\begin{equation}\label{eqReac2}
a/2[011](11\overline{1})\rightarrow  a/2[\overline{1}01](111)+ a/2[110](001).
\end{equation}

To the best of our knowledge, only one atomistic simulation of the nucleation of a 90$\degree$ MD through the induced nucleation of a complementary 60$\degree$ MD has been presented so far \cite{Ichimuraa1995}. However, a simplified quasi 
two-dimensional model of the Ge/Si(100) system was used and surface reconstruction was 
not included, so the activation energy for nucleation, for example, is not expected to be accurately estimated.
To better understand the process of the 90$\degree$ MD formation, one has to find minimum energy paths 
for this transition on the energy surface in a large enough atomic scale representation of the system.
This is a nontrivial problem due to the high dimensionality of the configuration space.   

We have recently explored the atomic relaxation mechanisms at the
microscopic scale in three-dimensional heteroepitaxial systems with hexagonal symmetry \cite{Trushin2009}.
Starting from the initial epitaxial state, we generated the final configuration containing various
kinds of defects with the repulsive bias potential method (RBP) \cite{Trushin2004}. 
Then, the nudged elastic band method \cite{Jonsson98,Jonsson2011} was employed to determine a minimum
energy path for the transition. This 
approach allowed us to 
classify the atomic mechanisms leading to strain relaxation in two
and three dimensional systems.

In the present work we apply
this procedure to a more complex system, an
atomistic model of the Ge/Si(100) system including the $(2 \times 1)$ surface reconstruction.
We note that under experimental conditions where defects and impurities are present,
the nucleation of dislocations may not occur through the homogenous nucleation process.
Due to lack of detailed experimental data realistic modeling of such processes is often challenging.
In the present work we consider homogenous nucleation only for the sake of conceptual clarity.
A minimum energy path for the formation
of a 90$\degree$ MD is found.
A possible mechanism is identified that starts with the nucleation of a 60$\degree$ MD. 
Then, a complementary 60$\degree$  MD nucleates with 
a 
lower activation energy and the two MDs 
finally
react to form the 90$\degree$  MD. 
While the activation energy is estimated to be very high, too high to be overcome by thermal 
activation alone, the mechanism identified illustrates how the presence of one MD can induce the formation of a 
second one. The origin of the high activation energy is probably related to limitations 
in the model, such as the form of the interaction potential
used and the absence of defects in the initial, 
epitaxial state. Further work is needed to resolve these issues.

\section{Model}

\begin{figure}
\includegraphics[width=\textwidth]{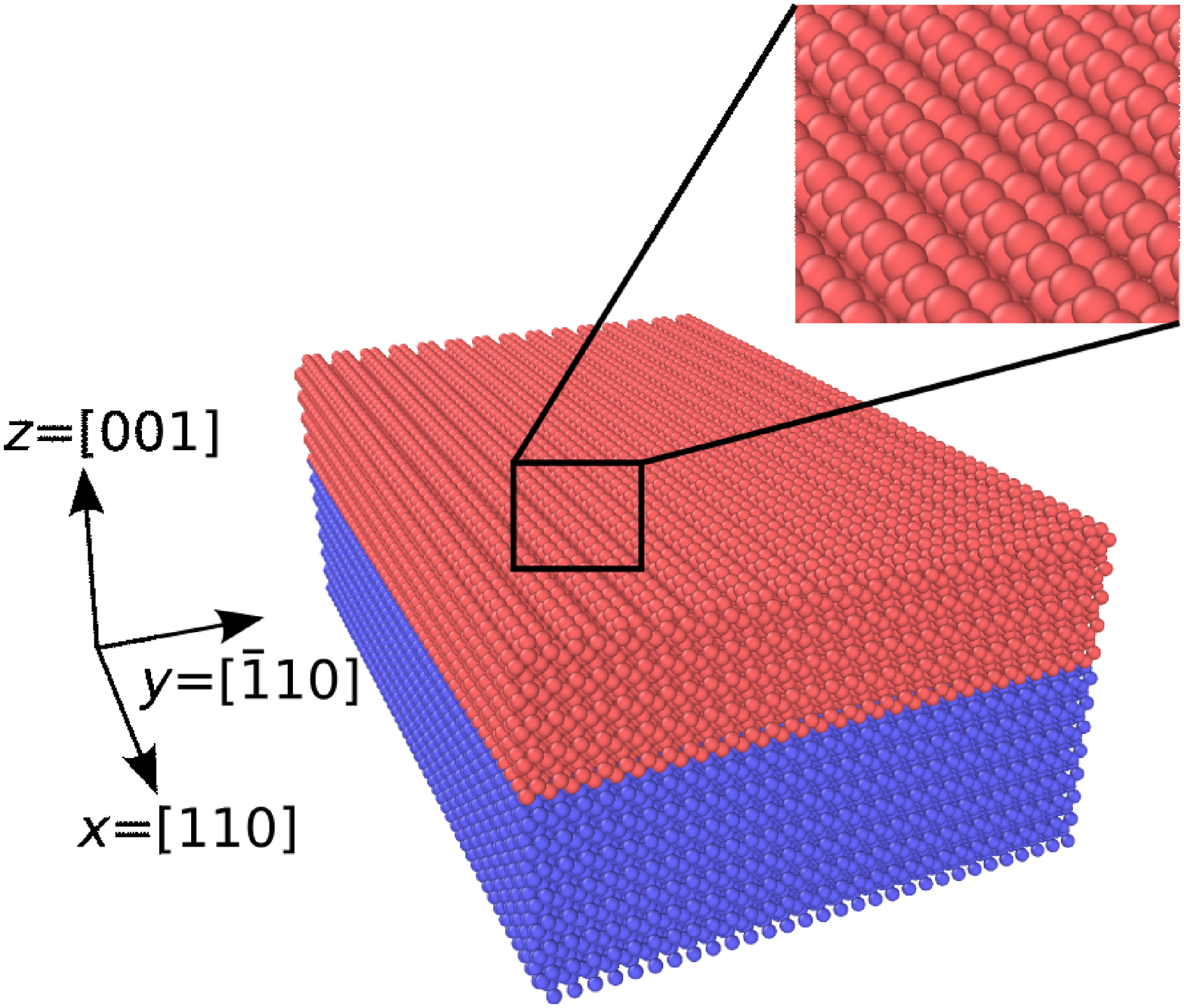}
\caption{ The model of the Ge/Si(001) system in the inital, coherent state. Blue spheres represent the Si  atoms and red represent the Ge atoms.  The surface is reconstructed as rows of dimer form between surface atoms.
  } \label{fig1}
\end{figure}

The computational model system has the shape of a parallelepiped, as shown in Fig. \ref{fig1},
with periodic boundary conditions applied in the 
$x-y$ plane to mimic an extended slab consisting of $31$ atomic layers of Si (representing the substrate), and $19$ layers of Ge (representing the film). 
The crystal orientation of the surface of the substrate is $(001)$. 
The $x$, $y$ and $z$ axes are oriented along the  $[110]$ , $[\overline 110]$, and $[001]$  directions, respectively. 
The bottom two layers of the substrate were fixed to mimic interaction with an extended Si crystal below the slab 
and to prevent the system from moving as a whole. 

Since dislocations create long distance strain fields, it is important to use 
as large a system as possible to minimize finite-size effects.  We checked that the system is large 
enough in the sense that further increase in the substrate thickness or in the system size in 
the $x$ and the $y$ directions does not significantly change the energetics of the transition paths
calculated here.   
Here, the aim is to generate dislocation lines oriented along the $y$ axis ($[\overline110]$ direction).
Because of periodic boundary conditions, a single MD 
in the computational cell corresponds to a network of parallel MDs. From experimental observations, 
the average distance between such MDs has been measured to be around $100$ \AA$ $  \cite{Liu2012}. 
Therefore, we chose the size of the system in the $x$ direction to accomodate $50$ atomic rows.
The lateral dimension of the system (in the $x-y$ plane) is thus $50 \times 30$ atomic rows, corresponding 
to a system of size of $19.2 \times 11.5 \times 6.8 $ nm$^3$. The total number of atoms in the model is $75000$.

The canonical Stillinger-Weber (SW) potential 
function is used to describe the interatomic forces \cite{Stillinger1985}. The SW potential is, 
of course, only a crude approximation to the atomic interactions, 
but it is the most widely used potential function and we have therefore 
adopted it here. The system size needed to describe the phenomena studied here 
makes the computational effort in quantum mechanical density functional calculations prohibitively large.   
The SW potential enforces sp$^3$ hybridization while in reality the bonding arrangements 
are more flexible \cite{Smith1995}. As a consequence, the results presented here are 
expected to give an overestimate of the activation energy for the dislocation formation.
The (001) surface is dimer reconstructed (cf. Fig. 1)
in the calculations,
as has been observed during
growth of Ge films on Si(001) \cite{Voigtlaender1999}.
Experimental observations, furthermore, show a complex $2\times N$ reconstruction pattern on the Ge/Si(001) surface where every $N$-th dimer is missing.
We used the ideal ($2\times 1$) reconstruction as an initial state in our model for the sake of simplicity.
The parameters of the Stillinger-Weber potential are given in Table \ref{table1}. They are the same as the one used in \cite{Ichimuraa1995,Laradji1995}.

\begin{table}
\centering
\caption{\label{table1}Parameters of the Stillinger-Weber potential for mixed Ge-Si interactions,taken from ref. \cite{Laradji1995}.}
{\begin{tabular}{@{}cccc}

\br
Parameter &  Si-Si & Ge-Si & Ge-Ge \\
\mr
$\sigma$    &   2.0916431      &   2.1354      &  2.17912051     \\
$\epsilon$  &  2.17  & 2.0427  &  1.93 \\
$\lambda$   &  21.0  &     -    &  31.0\\
\br
\end{tabular}}
\end{table}

\section{Method}

Epitaxial films with thickness beyond the equilibrium critical thickness will tend to
relax to a state that includes defects, 
but this process typically 
inolves overcoming an energy barrer $\Delta E$ during the transition from the
coherent state.
To find the activation energy and determine the mechanism of the transition,
a minimum energy path for the transition needs to be found. 
A direct simulation of the classical dynamics of the atoms is typically not useful
because the transition is a rare event on the time scale of the atomic vibrations.
While the transition rate can be increased by increasing the temperature,
a cross-over to a different mechanism, favored by entropy, is likely to occur. 
To clarify the mechanism of the relaxation, the identification of a minimum energy path is more useful.
However, this is a challenging task as many atoms are involved and it is not clear what the atomic structure of the final, relaxed state is.   

To generate an isolated defect at a given location in the sample, we have previously
introduced \cite{Trushin2004} a particularly simple but
efficient method called the repulsive bias potential (RBP) method.
In the RBP method, the system is
placed in a fixed, external repulsive bias potential which makes the
initial state unstable 
\begin{equation}
U_{\rm tot}(\vec r,\vec r_0)=U(\vec r)+A\exp\{-[(\vec r-\vec
r_{0})/\alpha]^2\}.
\label{rbp_potential}
\end{equation}
Here the components of $\vec{r}_0$ and $\vec{r}$ contain all the
atom
coordinates of the initial and current configurations, respectively.
The $U(\vec r)$ term contains the atomic interactions (here the Stillinger-Weber potential function)
and the RPB potential is the additional, spherically symmetric Gaussian term 
with strength $A$, range $\alpha$ and
a maximum value at $\vec r_{0}$. When $A$ and $\alpha$ have been chosen
appropriately, forces computed from Eq. (\ref{rbp_potential}) can
displace the system from its initial state 
to a nearby local minimum on the energy surface.
In practice, this is done
by applying an energy minimization using $U_{\rm tot}$ as objective function.

First, the initial epitaxial state is prepared by minimizing the total energy.
We have used a procedure based on classical dynamics where
the velocity of each atom is set to zero 
whenever it has a component opposite to the direction of the
acceleration \cite{Jonsson98,Jonsson2011}. 
The standard leap-frog algorithm was used to
numerically integrate the equations of motion.
The minimization was considered to be converged when the 
maximal force acting on an atom had dropped to less than $10^{-4}$ eV/\AA. 

To generate an atomic configuration for the final, defected state, the RBP bias is turned on and 
a group of atoms is displaced from the initial
position to bring the system closer to the 
final state of the
particular relaxation process we want to study. 
Rather than trying random initial displacements, some knowledge of the
defect generation mechanism is useful for expediting the process.
Then, the total energy minimization is reapplied to determine the structure at the new, defected local minimum.

It is important to note that this method can generate many
different final states depending on both the choice of initial displacements
and the exact form of the RBP. By making the
repulsive bias sufficiently localized around the initial potential
minimum, the final state energy depends only on the true potential
of the system and not on the fictitious repulsive bias.
We consider only final configurations with
precisely one isolated defect, namely a 90$\degree$  MD
to unravel the mechanism and energetics of its formation.  

To generate an isolated 90$\degree$  MD at certain position in the system, one
needs to use preliminary knowledge about the geometry of the diamond lattice and the arrangement of atoms in the dislocation core to select particular set of displacements and parameters of the RBP.  
Our trial and error attempts resulted in the following procedure:
Atoms in a wedge-shaped triangular prismoid
were displaced by $1.5$ \AA \ in the direction of $[\overline{1}11]$. 
The RBP parameters were chosen to be
$A = 2000 $ eV and $\alpha = 0.01 1$/\AA$^2$.
The minimization of $U_{\rm tot}$ then suffices to
move the system away from the potential basin corresponding to the perfect, epitaxial state to the 
energy minimum corresponding to a 90$\degree$  MD state. 

With both an initial and final state atomic configuration available, the nudged elastic band (NEB) method \cite{Jonsson98,Jonsson2011}, 
is used to find a
minimum energy path (MEP)
between the two.
A set of intermediate configurations ('images' of the system) are first generated to 
create a discretized path between the initial and final states. 
Iterative displacements of the images along the NEB force using some minimization algorithm then gradually bring the images to an MEP. 
When more than one MEP on the energy surface connects the given endpoints, the NEB minimization typically converges to the MEP closest to the initial path.
The initial location of the images is often obtained by
linear interpolation of the atomic coordinates between the initial
and final states.  
For the present calculations however, we found that this can lead to numerical instabilities due to the
strong hard core repulsion of the interatomic potentials as atoms are brought close to each other.
A method involving interpolation of pairwise distances has recently been presented to circumvent this problem \cite{Smidstrup2014}.
Here, however, we used the intermediate configurations obtained during the minimization with the repulsive bias. 
This approach leads to fast convergence of the NEB without the instabilities encountered with the linear interpolation scheme.
The same procedure has previously been used to study MD nucleation in metal-on-metal
systems with the FCC(111) surface orientation \cite{Trushin2009}.
In the NEB optimization, we use the same procedure based on classical dynamics with velocity zeroing as for the RBP minimization  \cite{Jonsson98,Jonsson2011}. 
To get reasonable resolution of the transition path, we used about 130 images in the NEB.


\section{Results }

We will first discuss the influence of the film thickness on the energy of the states with either a 
60$\degree$ MD or a 90$\degree$ MD as compared to the 
energy of the coherent epitaxial state. Then the MEP found for the formation of the 90$\degree$ MD is described.

\subsection{Relaxation energy}

\begin{figure}
\includegraphics[width=\textwidth]{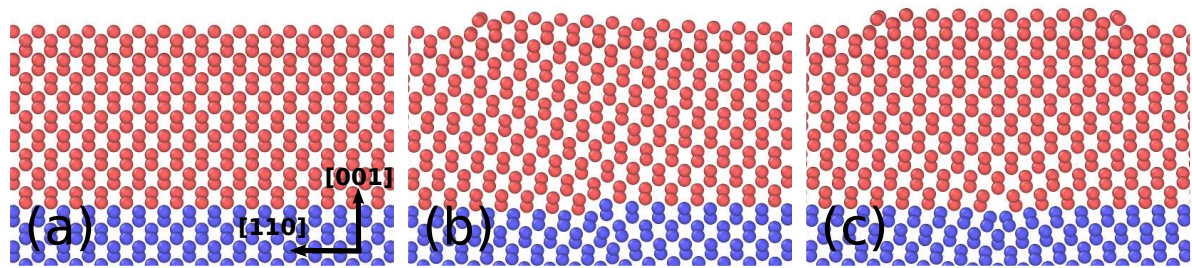}
\caption{  (a) Side view of the initial, coherent state. (b) Side view of the straight 60$\degree$ misfit dislocation. (c) Side view of the straight 90$\degree$ misfit dislocation. Blue spheres represent the  Si  atoms and red spheres the  Ge atoms.
  } \label{SideView}
\end{figure}

\begin{figure}
\includegraphics[width=\textwidth]{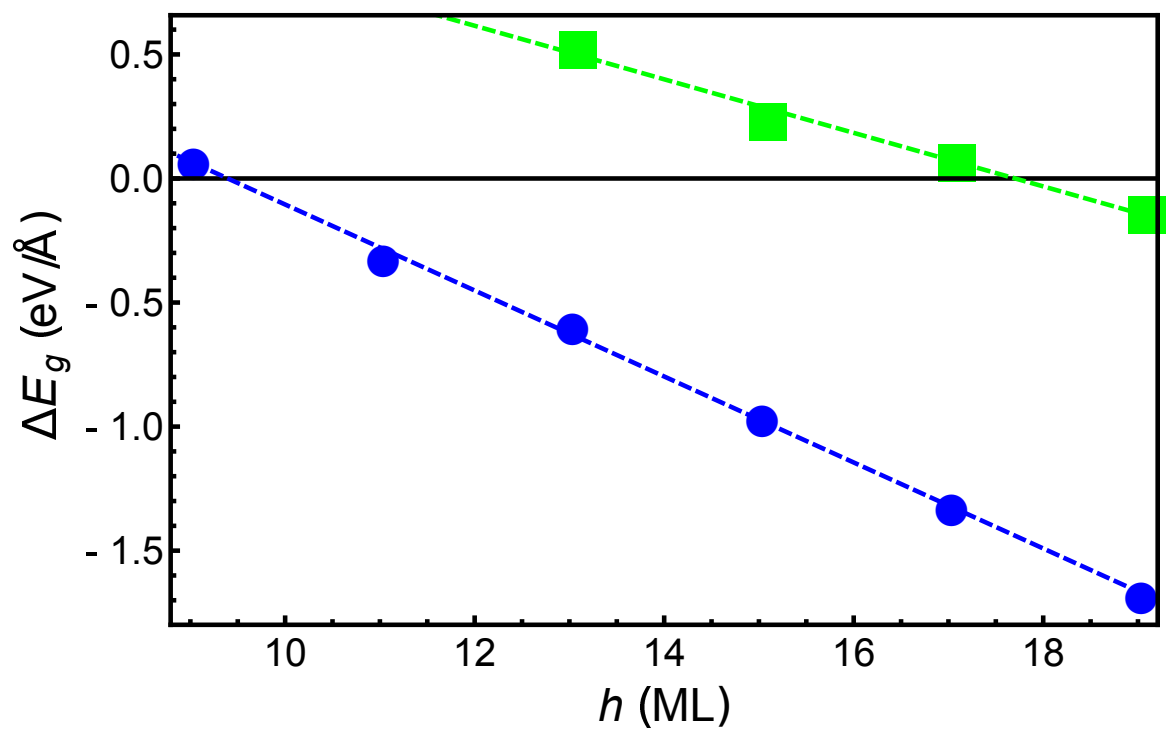}
\caption{Energy difference due to formation of a 60$\degree$  (green square) and a 90$\degree$ (blue dot) misfit dislocation (normalized to the length of the dislocation line)
 as function of the Ge film thickness. 
Dotted lines show a linear fit to the data. }
  \label{Gain}
\end{figure}

Starting from the coherent epitaxial state (Fig. \ref{SideView}a), the formation of a 60$\degree$ MD (Fig. \ref{SideView}b) or a 90$\degree$ MD (Fig. \ref{SideView}c) decreases the film strain by moving atoms out of the surface layer to form an island of adatoms. For a 90$\degree$ dislocation, the island is flat and its width is proportional to the film thickness. 
The energy difference $\Delta E_g = E_2 - E_1$ between the coherent, epitaxial state and a state where a MD has formed arises from competition between decrease in energy due to strain release in the film and increase in energy associated with the formation of the dislocation core and surface defects.
To study this, the Ge film thickness was varied between 9 and 19 ML 
while keeping 
31 atomic layers in the Si substrate.
Fig. \ref{Gain}  shows how $\Delta E_{g}$ corresponding to the  
formation of a 60$\degree$ MD and of a 90$\degree$ MD vary approximately linearly with the thickness of the Ge film.
This linear behavior arises from the fact that the decrease in strain energy is almost proportional to the film thickness while the energy of dislocation core and surface defects energy is independent of the film thickness for films thicker than a few MLs.
The results show that the energy of the system is lowered by forming a 60$\degree$ MD beyond
a critical film thickness of 19 atomic layers, which 
corresponds to approximately 2.7 nm.
This estimate agrees well with the results of Ichimura {\it et al.} \cite{Ichimuraa1995}.
The energy of the system is lowered by forming a 90$\degree$ MD beyond a critical film thickness of 
10 atomic layers, which corresponds approximately to 1.5 nm. 
This estimate agrees well with experimental observations \cite{Bevk1986}, 
but is a little higher than that obtained by of Ichimura {\it et al.}, 0.8 nm  \cite{Ichimuraa1995}.


\subsection{Minimum energy path}

\begin{figure}
\includegraphics[width=\textwidth]{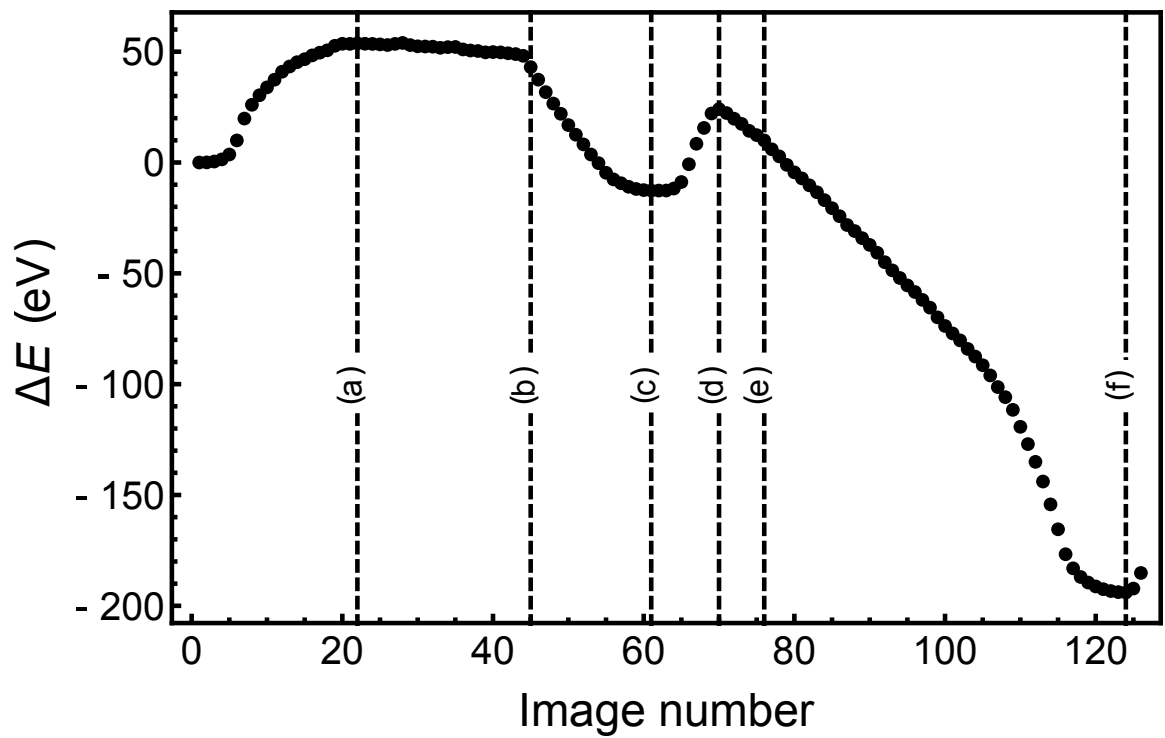}
\caption{ Energy along the minimum energy path between a defect free, coherent initial state 
and a final state with a single 90$\degree$  dislocation at the Ge/Si interface. The configurations corresponding to the dashed vertical lines 
are shown in Figs. \ref{Mechanism} and \ref{MechanismSchem}. The Ge film contains 19 atomic layers while the Si(001) substrate contains
31 atomic layers.
 } \label{Profile90}
\end{figure}

\begin{figure}
\includegraphics[width=\textwidth]{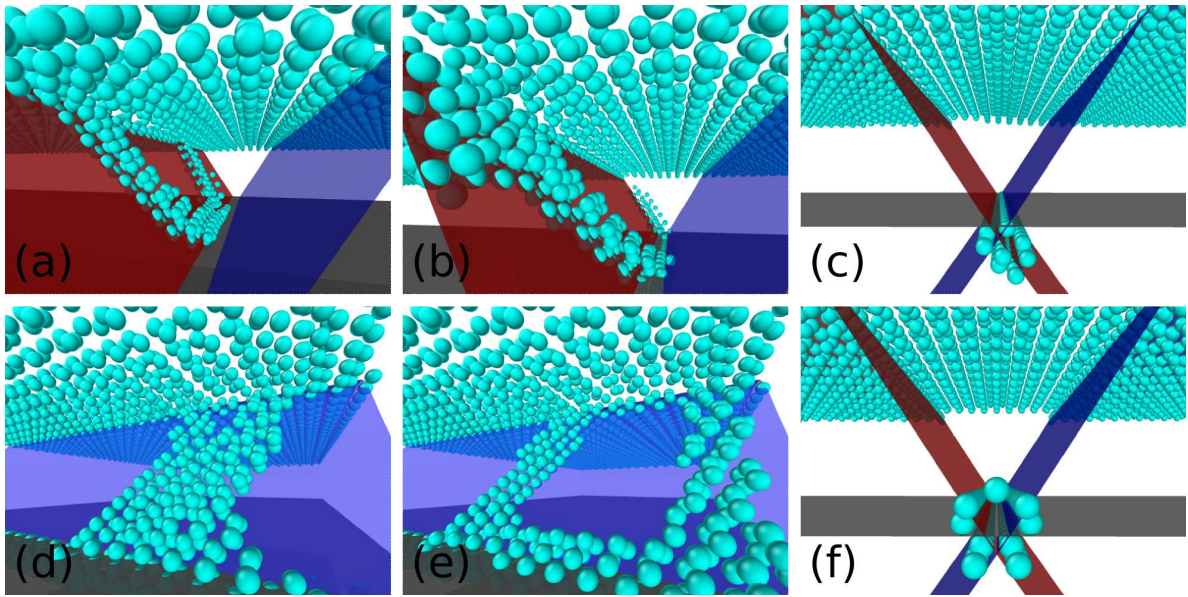}
\caption{ Six intermediate configurations along the minimum energy path for the formation of a 
90$\degree$ dislocation at the Ge/Si interface. The position of the configurations in the energy profile is indicated by vertical dashed lines on Fig. \ref{Profile90}. (a) corresponds to the first saddle point, (b) corresponds to the closing of the half-loop due to periodic boundary conditions, (c) corresponds to a straight 60$\degree$  dislocation, (d) corresponds to the second saddle point, (e) illustrates the growth of the 90$\degree$ dislocation, and (f) corresponds the straight 90$\degree$ dislocation. Only atoms whose surrounding does not correspond to a diamond lattice according to a common neighbor analysis \cite{Honeycutt1987,Clarke1993} as implemented in OVITO are shown \cite{Stukowski2010a}. The $(11\overline{1})$ and $(111)$ planes are shown in red and blue. The $(11\overline{1})$ plane is not shown in figures (d) and (e). 
The location of the Ge-Si interface is indicated by the gray plane.
 } \label{Mechanism}
\end{figure}

\begin{figure}
\includegraphics[width=\textwidth]{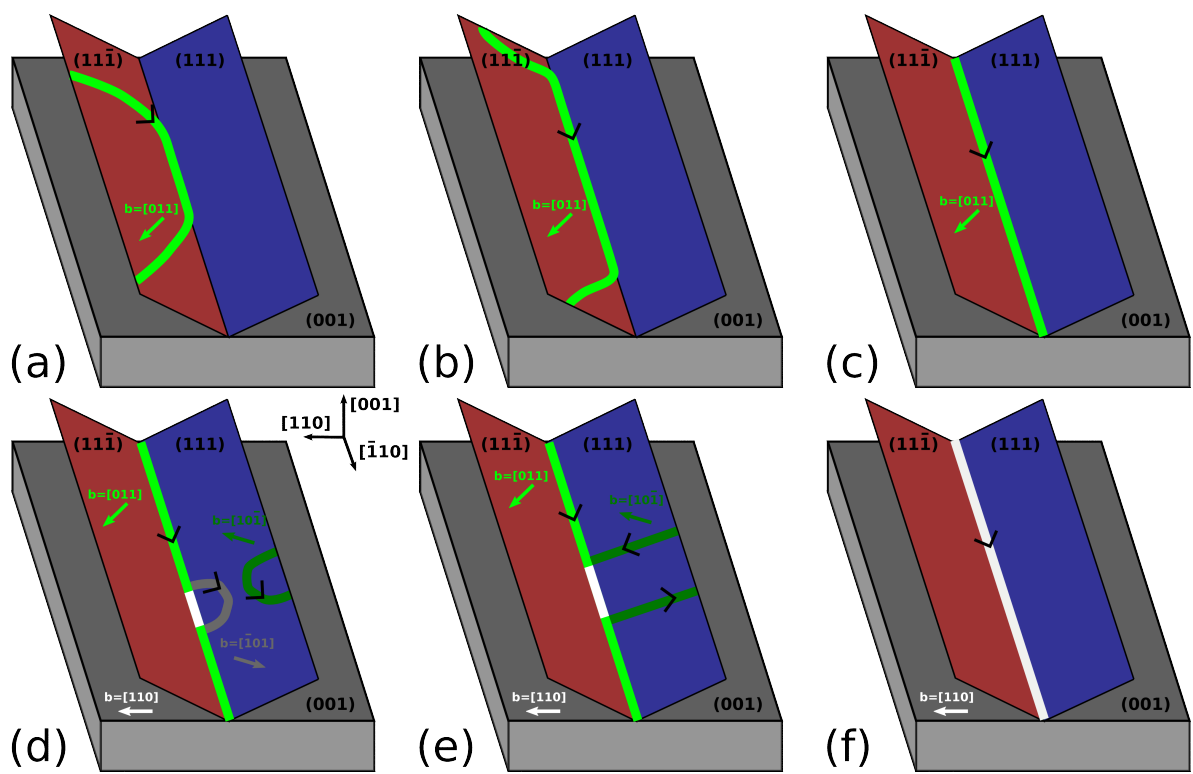}
\caption{ Schematic representation of the configurations shown in Fig. \ref{Mechanism} illustrating the mechanism for the formation of 
90$\degree$ misfit dislocation at the Ge/Si interface. 
The light green, dark green and gray lines indicate 60$\degree$ dislocations with Burgers vector $[011]$, $[10\overline{1}]$ and $[\overline{1}01]$, respectively. The white line indicates a 90$\degree$ misfit dislocation with Burgers vector $[110]$. The $(11\overline{1})$ and $(111)$ planes are shown in red and blue.
 } \label{MechanismSchem}
\end{figure}

The energy along the MEP found for the formation of  
a single, straight $90^{\degree}$ MD at the interface starting from the coherent epitaxial state is shown in Fig. \ref{Profile90}. A few 
intermediate configurations along the reaction path are shown in Figs. \ref{Mechanism} and \ref{MechanismSchem}.
A video can be downloaded from the supplemental material \footnote{\url{http://figshare.com/articles/Movie_ogg/1423292}}.


The reaction can be divided into two parts. The first part corresponds to the formation of a 60$\degree$  
MD with Burgers vector $[011]$ 
and has an activation energy of 54 eV. The Burgers vector was determined using the dislocation extraction 
algorithm \cite{Stukowski2010}. 
Figures \ref{Mechanism}a and \ref{MechanismSchem}a represent the configuration of the saddle point for the 
formation of the 60$\degree$  MD. 
They clearly show that the 60$\degree$  MD forms through a half-loop nucleation on the $(11\overline{1})$ slip plane. 
At the saddle point the dislocation has reached the interface and can be described as a straight dislocation 
at the interface terminated by two threading arms reaching the surface. 
The dislocation then spreads by glide of the two threading arms
in opposite directions. This dislocation growth is associated with a decrease of the energy due to the 
strain release in the film. 
The energy decrease is small in this case
because the film thickness corresponds to the critical thickness for the formation of a 60$\degree$  MD. 
Due to the periodic boundary conditions, the threading arms meet (see Figs. \ref{Mechanism}b and \ref{MechanismSchem}b) 
and annihilate each other leaving only a straight 60$\degree$  MD at the Ge/Si interface (shown in Figs. \ref{Mechanism}c and \ref{MechanismSchem}c). 
This annihilation leads to a decrease of the dislocation length and a decrease in energy. 
The straight 60$\degree$  MD corresponds to the intermediate local minimum in the reaction path and has slightly lower energy than the coherent state. 
Figure \ref{SideView}b shows that the formation of the 60$\degree$  MD also induces the formation of a double layer step on the surface.

In the second part of the reaction, a complementary 60$\degree$ MD with Burgers vector $[10\overline{1}]$ 
forms in the (111) plane which is the mirror plane of the $(11\overline{1})$ 
\footnote{According to \cite{Marzegalli2013}, the complementary MD is more likely to form not on the exact mirror plane 
but on a slightly translated mirror plane. After a glide of the complementary MD,  a pair of coupled 60$\degree$  
MDs can be observed near the interface \cite{Vila1996,Stirman2004}. 
In our calculations, a 90$\degree$  MD is present in the final state and the complementary MD thus has to form on the mirror plane.}. 
The activation energy for the formation of this complementary MD is around 37 eV,  30\% lower than the activation energy for the formation 
of the first 60$\degree$  MD. The presence of the first MD thus assists the formation of the complementary MD on the mirror plane.

In previous studies, it has been 
assumed that the complementary MD nucleates either from the free surface \cite{Marzegalli2013} or from the existing MD \cite{Gosling1993}.
In our calculations, both events occur simultaneously as one loop forms from the existing MD while a half-loop forms from the surface
(see Figs. \ref{Mechanism}d and \ref{MechanismSchem}d). 
The two loops then merge to leave only the threading arms shown in Figs.  \ref{Mechanism}e and \ref{MechanismSchem}e. 
The threading arms can then glide and react with the straight 60$\degree$  MD to form a 90$\degree$  MD according to Eq. \ref{eqReac}. 

The final state of the system with a single 90$\degree$  dislocation is shown in Figs. \ref{Mechanism}f,  \ref{MechanismSchem}f  and \ref{SideView}c. 
It is characterized by the appearance of an extended double layer island on the surface running along the $y$ direction.


\section{Discussion}

The estimate of the activation energy obtained here, 54 eV, is obviously very high. 
An event with such a high activation energy would never occur by thermal fluctuations even 
if the temperature were close to the melting temperature. 
However, it has been experimentally shown that the half-loop nucleation of a 60$\degree$ dislocation can 
occur even at a lower strain in a Ge$_{0.32}$Si$_{0.68}$/Si(001) film\cite{Bolkhovityanov2004}. 
The mechanism presented here 
might therefore be possible. This implies that we are significantly overestimating the activation energy in our calculations. 
Unfortunately no reliable experimental estimate of the activation energy is available in the literature.

Many atomistic calculations of dislocation nucleation in semi-conductor materials have predicted 
a very high activation energy \cite{Bolkhovityanov2001,Ichimuraa1995,Hull1989,Li2010}.
Below, we discuss possible explanations for these large values. 

\subsection{Accuracy of the potential}

The SW potential enforces 
chemical bonding according to sp$^3$ hybridization 
while in reality the bonding arrangements can be more flexible \cite{Smith1995}. The chemical bonding between
atoms in the core of the dislocation is, in particular, expected to deviate from sp$^3$ hybridization and
thereby be overestimated in the present simulations.
In order to obtain an estimate of this effect, the difference in energy per atom was calculated for each atom 
to identify which region of the dislocation is primarily responsible for the activation energy of the formation of a 60$\degree$ MD. The total increase in energy of the core atoms was found to be 61.7 eV. 
If the energy of core atoms is overestimated by 30\%, the activation energy would be overestimated by around 20 eV. 
In principle, our calculations could be corrected by estimating the error in the dislocation core energy from comparison with more accurate density function theory (DFT) calculations. However, estimating dislocation core energy from DFT is a challenging task \cite{Bacon2009,Pizzagalli2003,Clouet2009} and to the best of our knowledge no estimate of dislocation core energy in Ge is available.

Preliminary calculations have also been carried out using a different potential function, 
the Tersoff potential \cite{Tersoff1988,Tersoff1989,Tersoff1990}, and the 
activation energy is then found to be 12 eV lower than with the SW potential. 
The large difference in the activation energy calculated with 
these two potential functions indicates that a more accurate description of the atomic interactions allowing for 
deviations from sp$^3$ hybridization could 
give a significantly lower activation energy.
Other potentials such as the Lenosky potential \cite{Lenosky2000} have been shown to 
more accurately decribe dislocation cores in Si \cite{Pedersen2009} but a parametrization for Ge is not available.
There is clearly a need for developing a more accurate potential function 
for modeling dislocation cores in the
Ge/Si system.

\subsection{Absence of defects}

\begin{figure}
\includegraphics[width=\textwidth]{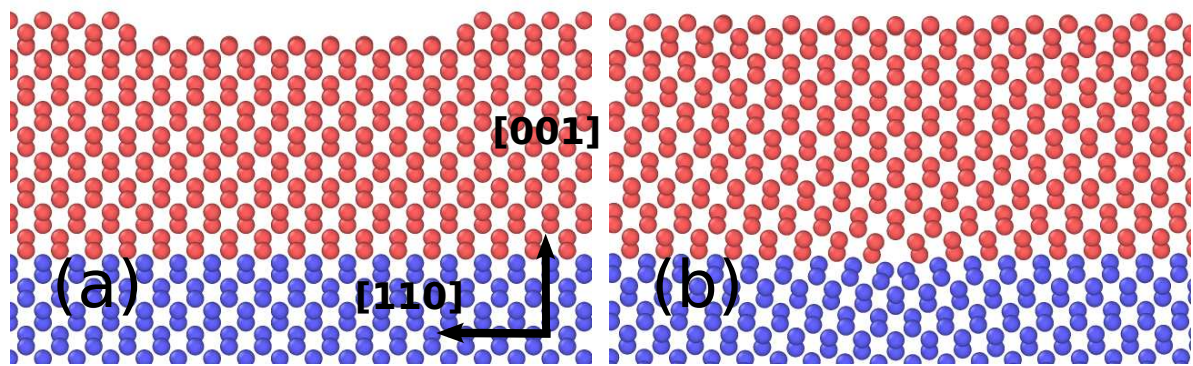}
\caption{ Side view of two configurations for a Ge/Si(001) film. Starting from a film containing steps (a), the formation of a straight 90$\degree$ MD leads to the elimination of the steps (b).} \label{SideViewStepInfl}
\end{figure}

Dislocations usually nucleate from pre-existing defects. For instance, it was  found experimentally 
that most of the 60$\degree$ MD half-loops nucleated from the same location in a Ge$_{0.32}$Si$_{0.68}$/Si(001) film \cite{Bolkhovityanov2004}. 
This observation was interpreted as indicating that a
defect was initiating the formation of the dislocation. Such defects can be impurities \cite{Barnoush2010,Hull1989,Li2012} or, more importantly, steps on the surface. 
When starting from a defect free film, the formation of MDs leads to the formation of double-layer steps on the surface 
as shown in Fig. \ref{SideView}. This is energetically unfavorable since step atoms have dangling bonds. 
At the saddle point the increase in energy due to the step edge atoms is 7.7 eV. 
If the film initially contains steps as indicated in Fig. \ref{SideViewStepInfl}(a), the formation of an MD eliminates the surface steps, as illustrated in Fig. \ref{SideViewStepInfl}(b). The activation energy for forming an MD would, thereby,
be lowered by the presence of steps in the initial state. Steps, furthermore, act as stress concentrators and most of the atomic scale simulations of the formation of 60$\degree$ MD in diamond systems have, indeed, been carried out for systems that include a surface step in the initial state \cite{Godet2004,Godet2006,Godet2009,Shima2010,Li2010,Li2012,Izumi2008,Marzegalli2005,Marzegalli2005a}.

Since at the saddle point, the increase in energy due to the step edge atoms is 7.7 eV, 
we can expect that having a straight step in the initial configuration 
would lower the activation energy by about 15 eV. The lowering of the activation energy of 
the step should be similar for the nucleation of the initial and of the 
complementary MD but a significant difference should be noted. Formation of an initial 
MD will be favored in the vicinity of a step whereas in order to favor the formation of a 
complementary MD a step has to be located on the mirror (111) plane of an 
existing 60$\degree$ MD. 
During growth of the film, it can be expected that steps are propagating 
along the surface
as atoms get deposited on the surface and that a step will at some time be located 
at just the right position for inducing the nucleation of the complementary MD.

\section{Summary and Conclusions}

Misfit strain relaxation in the heteroepitaxial Ge/Si(001) system has been studied using
an atomistic simulation method where 
a repulsive bias potential is used  
to generate atomic coordinates of a 90$\degree$ MD
and a nudged elastic band calculation is then carried to to find a minimum energy 
path between this and the coherent epitaxial state.
The Stillinger-Weber interatomic potential energy function has been used to 
approximate the atomic interactions. 
The minimum energy path found here for the 
formation of the 90$\degree$ MD 
involves
half-loop nucleation of a 60$\degree$  MD and the subsequent induced nucleation of a 
complementary 60$\degree$  MD. The activation energy for the formation of the first and 
second 60$\degree$  MD is 54 eV and 37 eV, respectively. The presence of the first MD thus assists
the formation of the second by lowering the activation energy by 30\%.
The large value of the activation energy obtained here may be attributed to the absence of 
defects in the initial state of the film and/or to the inflexibility of the Stillinger-Weber 
potential function which does not account for bonding 
arrangements that deviate from sp$^3$ hybridization and thus gives very high energy for the 
atoms in the dislocation core of a 60$\degree$  MD.


\section{Acknowledgments}

This work was supported by the Academy of Finland through the FiDiPro program 
(H. J., grant no. 263294) and the COMP CoE (T. A-N, grant no. 251748). 
We acknowledge computational resources provided by the Aalto Science-IT project and CSC IT Center for Science Ltd in Espoo, Finland.
E.G. was supported by Funda\c c\~ao de Amparo
\`a Pesquisa do Estado de S\~ao Paulo - FAPESP (Grant no.
 2014/15372-3). O.T. was supported by Russian Foundation for Basic Reserch 
grant No. 14-00139a. S.C.Y was supported by a Brazilian Initiative Collaboration Grant funded by the Watson Institute at Brown University. E.M. wish to thank Laurent Pizzagalli for helpful discussions.


\bibliographystyle{tPHM}
\bibliography{Biblio.bib}


\end{document}